\documentclass[a4paper,12pt]{article}

\begin{document}
\makeatletter
\renewcommand{\theequation}{\thesection.\arabic{equation}}
\@addtoreset{equation}{section}
\makeatother

\title{Holographic Relation in Yang's Quantized Space-Time Algebra and Area-Entropy Relation in $D_0$ Brane 
Gas System. II}
 \author{\large Sho Tanaka\footnote{Em. Professor of Kyoto University, E-mail: st-desc@kyoto.zaq.ne.jp }
\\[8 pt]
 Kurodani 33-4, Sakyo-ku, Kyoto 606-8331, Japan}
\date{}
\maketitle

\vspace{-10cm}
\rightline{}
\vspace{12cm}
\thispagestyle{empty}

We investigate a possible inner relation between the familiar Bekenstein-Hawking area-entropy 
relation and ours presented in a simple $D_0$ brane gas model on the basis of the kinematical holographic 
relation [KHR] in the Yang's quantized space-time algebra (YSTA). We find out that the relation between them 
is well understood through the idea of an {\it elementary} Schwarzschild black hole realized on a single $[site]$ 
of Planckian scale in our scheme. Related arguments possibly explain the origin of a certain kind of universality 
as seen in $\eta =1/4$ in the Bekenstein-Hawking relation and lead us to notice the similarity between 
the effective mass of $D_0$ brane, $\mu_S,$ inside black holes and the Hawking radiation temperature, $T_{H.R.}
( = 1/(8\pi M_S) )$, both inversely proportional to the black hole mass $M_S$ and having almost the same order of 
magnitude. Motivated by this fact, we introduce in our scheme an ansatz which enables us self-consistently 
to equate $\mu_S$ to $T_{H.R.}$ and leads us to an area-entropy relation with $\eta$ slightly shifted from $1/4.$

\newpage
\section{\normalsize Introduction}

We put forward the idea presented in the preceding paper with the same title,$^{[1]}$ hereafter referred as $I$. In the 
latter paper, we have presented a new area-entropy relation [AER] in a simple $D_0$ brane gas model  on the basis 
of the kinematical holographic relation [KHR]$^{[2]}$ found in the Yang's quantized space-time algebra 
(YSTA),$^{[3],[1],[2]}$  and tried to make clear its possible inner relation to the ordinary Bekenstein-Hawking 
area-entropy relation.$^{[4]}$ 

Indeed, the arguments presented there is essentially based on the framework of [KHR], a kind of holographic relation in the Lorentz
-covariant Yang's quantized space-time algebra(YSTA), which we called the kinematical holographic 
relation (KHR).$^{[2]}$ It was simply expressed as
\begin{eqnarray} 
\hspace{-3cm} [KHR] \hspace{2cm}       n^L_{\rm dof} (V_d^L)= {\cal A} (V_d^L) / G_d,
\end{eqnarray}
that is, the proportional relation between $n_{\rm dof}(V_d^L)$ and ${\cal A} (V_d^L)$ with proportional constant $G_d$, 
where $n_{\rm dof}$ and ${\cal A}$, respectively, denote the number of degrees of freedom of any $d$ dimensional 
bounded spatial region with radius $L$, $V_d^L$, in Yang's quantized space-time and the boundary area of the latter region 
in unit of $\lambda$ or Planck length.

 As was emphasized in [2], the relation [KHR] essentially reflects the fundamental nature of the noncommutative 
geometry of YSTA, that is, a definite kinematical reduction of spatial degrees of freedom in comparison with 
the ordinary lattice space. It presents a possibility of giving a simple clue to resolve the long-pending problem 
encountered in the Bekenstein-Hawking area-entropy relation$^{[4]}$ or the holographic principle,$^{[5]}$ that is, 
the apparent gap between the degrees of freedom of any bounded spatial region associated with entropy and of its 
boundary area. 

As a matter of fact, we derived in I, ref. [1], a new area-entropy relation of the $D_0$ brane gas model$^{[6]}$ 
constructed on YSTA according to the idea of M-theory$^{[7]}$: 
\begin{eqnarray}
\hspace{-3cm} [AER] \hspace{2cm}  S(V_d^L) \leq {\cal A} (V_d^L) {S_S[site] \over G_d},
\end{eqnarray}
where  $S(V_d^L)$ and ${\cal A} (V_d^L)$ denotes the entropy of the gas system inside $V_d^L.$  One finds that 
the familiar Bekenstein proportional constant $\eta$ is now given by
 \begin{eqnarray}
\hspace{-3cm} \eta = {S_S[site] \over G_d},
\end{eqnarray}
where $S_S[site]$ denotes the entropy of individual [site], i.e., the fundamental constituent of $V_d^L$ with Planckian 
scale, hence the entropy of whole system being given by $S(V_d^L) = n_{dof} S[site].$  

As was pointed out in refs. [8], [2], [3], YSTA which is intrinsically equipped with short- and long-scale 
parameters, $\lambda (= l_P)$ and $R$, gives a finite number of spatial degrees of freedom for any finite spatial region 
and provides a basis for the field theory free from ultraviolet- and infrared-divergences. In this line of thought, 
in what follows, we will put forward the argument in I, as roughly explained above and make clear the possible inner 
relation between the familiar Bekenstein-Hawking area-entropy relation and our area-entropy relation, (1.2).

The present paper is organized as follows. In order to make the paper as self-contained as possible, we briefly 
review Yang's quantized space-time algebra (YSTA) and its representations$^{[8[,[2],[3]}$ in section 2. 
Section 3 is devoted to recapitulation of the kinematical holographic relation [KHR] ( in subsection 3.1) and to 
its extension to the lower-dimensional bounded regions, $V_d^L$ (in subsection 3.2). 
In section 4, subsections 4.1 
and 4.2 are devoted to review I, ref. [1]: In subsection 4.1, we first review a simple 
$D_0$ brane gas model and its statistical consideration through the entropy of the system constructed on the basic 
concept of [site] characteristic of noncommutative YSTA. In subsection 4.2, we find a new area-entropy 
relation in the system in connection with Schwarzschild black hole, as was shown in (1.2). 

In subsection 4.3, we newly present the basic arguments on the physical implication of Hawking radiation temperature 
$T_{H.R.} (= 1/(8\pi M_S)$ and Bekenstein-Hawking relation with $\eta =1/4$ in a close connection with our scheme. 
Based on these arguments, in the last section 5, we attempt to identify $S_S[site]$ in (1.3) with the entropy of 
an {\it elementary} Schwarzschild black hole realized on a single $[site]$ of Planckian scale and further find out 
an important fact that the effective mass of $D_0$ brane inside black holes, $\mu_S$, is inversely proportional to 
the black hole mass $M_S$, almost in accord with the Hawking radiation temperaturte, $1/(8\pi M_S).$ Motivatated by 
this fact, we introduce a new ansatz as the first principle in our scheme which enables us to equate self-consistently  
$\mu_S$ to $T_{H.R.}$ and leads us to an area-entropy relation with $\eta$ slightly shifted from $1/4.$

\section{\normalsize Yang's Quantized Space-Time Algebra (YSTA) and \break Its Representations}  

\subsection{\normalsize Yang's Quantized Space-Time Algebra (YSTA) }

Let us first recapitulate briefly the Lorentz-covariant Yang's quantized space-time algebra (YSTA).
$D$-dimensional Yang's quantized space-time algebra is introduced$^{[3],[8],[2]}$ as the 
result of the so-called Inonu-Wigner's contraction procedure with two contraction parameters, $R$ and 
$\lambda$, from $SO(D+1,1)$ algebra with generators $\hat{\Sigma}_{MN}$; 
\begin{eqnarray}
 \hat{\Sigma}_{MN}  \equiv i (q_M \partial /{\partial{q_N}}-q_N\partial/{\partial{q_M}}),
\end{eqnarray}
which work on $(D+2)$-dimensional parameter space  $q_M$ ($M= \mu,a,b)$ satisfying  
\begin{eqnarray}
             - q_0^2 + q_1^2 + \cdots + q_{D-1}^2 + q_a^2 + q_b^2 = R^2.
\end{eqnarray}
 
Here, $q_0 =-i q_D$ and $M = a, b$ denote two extra dimensions with space-like metric signature.

$D$-dimensional space-time and momentum operators, $\hat{X}_\mu$ and $\hat{P}_\mu$, 
with $\mu =1,2,\cdots,D,$ are defined in parallel by
\begin{eqnarray}
&&\hat{X}_\mu \equiv \lambda\ \hat{\Sigma}_{\mu a}
\\
&&\hat{P}_\mu \equiv \hbar /R \ \hat{\Sigma}_{\mu b},   
\end{eqnarray}
together with $D$-dimensional angular momentum operator $\hat{M}_{\mu \nu}$
\begin{eqnarray}
   \hat{M}_{\mu \nu} \equiv \hbar \hat{\Sigma}_{\mu \nu}
\end{eqnarray} 
and the so-called reciprocity operator
\begin{eqnarray}
    \hat{N}\equiv \lambda /R\ \hat{\Sigma}_{ab}.
\end{eqnarray}
Operators  $( \hat{X}_\mu, \hat{P}_\mu, \hat{M}_{\mu \nu}, \hat{N} )$ defined above 
satisfy the so-called contracted algebra of the original $SO(D+1,1)$, or Yang's 
space-time algebra (YSTA):
\begin{eqnarray}
&&[ \hat{X}_\mu, \hat{X}_\nu ] = - i \lambda^2/\hbar \hat{M}_{\mu \nu}
\\
&&[\hat{P}_\mu,\hat{P}_\nu ] = - i\hbar / R^2\ \hat{M}_{\mu \nu}
\\
&&[\hat{X}_\mu, \hat{P}_\nu ] = - i \hbar \hat{N} \delta_{\mu \nu}
\\
&&[ \hat{N}, \hat{X}_\mu ] = - i \lambda^2 /\hbar  \hat{P}_\mu
\\
&&[ \hat{N}, \hat{P}_\mu ] =  i \hbar/ R^2\ \hat{X}_\mu,
\end{eqnarray}
with familiar relations among ${\hat M}_{\mu \nu}$'s omitted.

\subsection{Quasi-Regular Representation of YSTA}

Let us further recapitulate briefly the representation$^{[8],[10]}$ of YSTA for the subsequent 
consideration in section 4. First, it is important to notice the following elementary fact that ${\hat\Sigma}_{MN}$ 
defined in Eq.(2.1) with $M, N$ being the same metric signature have discrete eigenvalues, i.e., $0,\pm 1 ,
\pm 2,\cdots$, and those with $M, N$ being opposite metric signature have continuous eigenvalues,
$\footnote{The corresponding eigenfunctions are explicitly given in ref. [10].}$ consistently with 
covariant commutation relations of YSTA. This fact was first emphasized by Yang$^{[3]}$ with respect to 
the pioneering Snyder's quantized space-time.$^{[9]}$ This conspicuous aspect is well understood by means of 
the familiar example of the three-dimensional angular momentum in quantum mechanics, where individual components, 
which are noncommutative among themselves, are able to have discrete eigenvalues, consistently with the 
three-dimensional rotation-invariance. 
 
This fact implies that Yang's space-time algebra (YSTA) presupposes for its representation space 
to take representation bases like 
\begin{eqnarray}
| t/\lambda,n_{12}, \cdots> \equiv |{\hat{\Sigma}}_{0a} =t/\lambda> |{\hat{\Sigma}}_{12}=n_{12}>
\cdots|{\hat{\Sigma}}_{910}=n_{910}>,
\end{eqnarray}
where $t$ denotes {\it time}, the continuous eigenvalue of $\hat{X}_0 \equiv \lambda\ \hat{\Sigma}_{0 a}$ 
and $n_{12}, \cdots$ discrete eigenvalues of maximal commuting set of subalgebra of $SO(D+1,1)$ which are 
commutative with ${\hat{\Sigma}}_{0a}$, for instance, ${\hat{\Sigma}}_{12}$, ${\hat{\Sigma}}_{34},\cdots , 
{\hat{\Sigma}}_{910}$, when $D=11$.$^{[2],[8],[10]}$

Indeed, an infinite dimensional linear space expanded by $|\ t/\lambda, 
n_{12},\cdots>$ mentioned above provides a representation space of unitary infinite dimensional representation of YSTA. It is 
the so-called "quasi-regular representation"$^{[11]}$ of SO(D+1,1),\footnote{It corresponds, in the case of unitary 
representation of Lorentz group $SO(3,1)$, to taking $K_3\ (\sim \Sigma_{03})$ and $J_3\ (\sim \Sigma_{12})$ to be 
diagonal, which have continuous and discrete eigenvalues, respectively, instead of ${\bf J}^2$ and $J_3$ in 
the familiar representation.}
and is decomposed into the infinite series of the ordinary unitary irreducible representations of 
$SO(D+1,1)$ constructed on its maximal compact subalgebra, $SO(D+1)$. 

It means that there holds the following form of decomposition theorem,
\begin{eqnarray}
| t/\lambda, n_{12},\cdots>= \sum_{\sigma 's}\ \sum_{l,m}\  C^{\sigma's, n_{12}, \cdots }_{l,m}(t/\lambda)\ 
 | \sigma 's ; l,m>,
\end{eqnarray}      
with expansion coefficients $C^{\sigma's, n_{12}, \cdots}_{l,m}(t/\lambda).^{[10], [8],[2]}$ In Eq.(2.13), 
$|\sigma 's ; l, m>'s$ on the right hand side describe the familiar unitary irreducible representation 
bases of $SO(D+1,1)$, which are designated by $\sigma 's$ and $(l,m),$  
\footnote{In the familiar unitary irreducible representation of $SO(3,1)$, it is well known that $\sigma$'s are 
represented by two parameters, $(j_0, \kappa)$, with $j_0$ being $1,2, \cdots \infty$ and $\kappa$ being purely 
imaginary number, for the so-called principal series of representation. With respect to the associated representation 
of $SO(3)$, when it is realized on $S^2$, as in the present case, $l$'s denote positive integers, 
$l= j_0, j_0+1, j_0+2,\cdots,\infty$, and $m$ ranges over $\pm l, \pm(l-1) , \cdots,\pm1, 0.$ } 
denoting, respectively, the irreducible unitary representations of $SO(D+1,1)$ and the associated 
irreducible representation bases of $SO(D+1)$, the maximal compact subalgebra of $SO(D+1,1)$, mentioned above. 
It should be noted here that, as remarked in [8], $l$'s  are limited to be integer, excluding the possibility of 
half-integer, because of the fact that generators of $SO(D+1)$ in YSTA are defined as differential operators 
on $S^D$, i.e., ${q_1}^2 + {q_2}^2 + \cdots + {q_{D-1}}^2 + {q_a}^2 + {q_b}^2 = 1.$

In what follows, let us call the infinite dimensional representation space introduced above for the representation of 
YSTA, Hilbert space I, in distinction to Hilbert space II which is Fock-space constructed dynamically by 
creation-annihilation operators of second-quantized fields on YSTA, such as $D_0$ brane field,$^{[10]}$ discussed 
in section 4.

\section{\normalsize Kinematical Holographic Relation [KHR] in YSTA}

\subsection{\normalsize Recapitulation of Kinematical Holographic Relation [KHR]} 

First, let us remember that the following kinematical holographic relation\footnote{The argument in this 
subsection was given in I $^{[1]}$, based on refs.[10], [2], in the following form of $D_0$ brane field equation: 
$[({X_\sigma}^2 +R^2 N^2 )( (\partial/\partial {X_\mu})^2 + R^{-2} (\partial/\partial {N})^2)) 
 - ( X_\mu \partial/\partial {X_\mu} + N \partial/\partial{N})^2 - (D-1)( X_\mu \partial/\partial {X_\mu}+N
 \partial/\partial{N})\ ]\ D ( X_\nu, N) = 0.$  Indeed, it was derived in ref.[10] from the following $D_0$ brane field 
action after M-theory,$^{[7]}$ 
$ \bar{\hat L} = A\ {\rm tr}\  \{ [\hat {\Sigma}_{KL}, 
\hat {D}^\dagger]\ [\hat {\Sigma}_{KL}, \hat {D}]\}=  A'\ {\rm tr}\  \{ 2\ (R^2 /\hbar^2)\  [{\hat P}_\mu, 
\hat {D}^\dagger ]\ [ \hat {P}_\mu, \hat {D}] - {\lambda}^{-4 }\ [\ [\hat {X}_\mu, \hat {X}_\nu], \hat {D}^\dagger] 
[ [\hat {X}_\mu,\hat {X}_\nu], \hat {D}]\},$ with $K, L = (\mu, b),$ by means of the Moyal star product method.} 
\begin{eqnarray}
\hspace{-3cm} [KHR] \hspace{2cm}       n^L_{\rm dof}= {\cal A} / G,
\end{eqnarray}
with the proportional constant $G$
\begin{eqnarray}
G\  \sim {(2 \pi)^{D/2} \over 2}\ (D-1)!! &&for\ D\ even
\\
    \sim  (2 \pi)^{(D-1)/2}(D-1)!! &&for\ D\ odd,
\end{eqnarray}
was derived in [2] for the $D$-dimensional space-like region with finite radius $L$ in D-dimensional 
Yang's quantized space-time in the unit of $\lambda$. Let us denote the region hereafter as $V_D^L$, 
which was defined by

\begin{eqnarray}
 \sum_{K \neq 0}{\Sigma_{aK}}^2  = \sum_{\mu \neq 0}{\Sigma_{a \mu}}^2 + {\Sigma_{ab}}^2 = (L/\lambda)^2,
\end{eqnarray}
or
\begin{eqnarray}
{X_1}^2 + {X_2}^2 + \cdots + {X_{D-1}}^2 + R^2\ N^2 = L^2.
\end{eqnarray}
Here, $\Sigma_{MN}$'s are presumed to be given in terms of Moyal star product formalism applied to the 
expression, $\Sigma_{MN}= ( -q_M p_n + q_N p_M)$, as was  treated in detail in [2].

${\cal A}$ in\ [KHR] (3.1) simply denotes the boundary surface area of $V_D^L$, that is,  
\begin{eqnarray}
{\cal A} =  ({\rm area\ of}\ S^{D-1}) ={(2 \pi)^{D/2} \over {(D-2)!!}} (L/\lambda)^{D-1} 
&&for\ D\ even,
\nonumber\\                   =2 {(2\pi)^{(D-1)/2} \over {(D-2)!!}} (L/\lambda)^{D-1} &&for\ D\ odd. 
\end{eqnarray}

On the other hand, $n^L_{\rm dof}$ in [KHR] (3.1), which denotes, by definition, the number of spatial 
degrees of freedom of YSTA inside $V_D^L$, was given in [2] as follows,  
\begin{eqnarray}
&&n^L_{\rm dof}  =  dim\ ( \rho_{[L/\lambda]}) \sim {2 \over (D-1)!}{([L/\lambda]+ D-2)! \over ([L/\lambda]-1)!}
\nonumber\\
 &&\hskip3.5cm \sim {2 \over (D-1)!}  [L/\lambda]^{D-1}.
\end{eqnarray}

Indeed, the derivation of the above equation (3.7) was the central task in [2]. In fact, we emphasized that 
the number of degrees of freedom $n^L_{\rm dof}$ inside $V_D^L$, which is subject to noncommutative algebra, YSTA, 
should be, logically and also practically, found in the structure of representation space of YSTA, that is, 
Hilbert space I defined in section 2. Let us here recapitulate in detail the essence of the derivation in order to 
make the present paper as self-contained as possible. 
   
In fact, one finds that the representation space needed to calculate $n^L_{\rm dof}$ is 
prepared in Eq.(2.13), where any "quasi-regular" representation basis,\\ $ | t/\lambda, n_{12}, \cdots>$, 
is decomposed into the infinite series of the ordinary unitary representation bases of 
$SO(D+1,1)$, $| \sigma 's ; l,m>.$ 
As was stated in subsection 2.2, the latter representation bases, $| \sigma 's ; l,m>'s$ are constructed on 
the familiar finite dimensional representations of maximal compact subalgebra of YSTA, $SO(D+1)$, whose representation 
bases are labeled by $(l,m)$ and provide the representation bases for spatial quantities under consideration, 
because $SO(D+1)$ just involves those spatial operators $( \hat{X}_u, R \hat{N})$. 

In order to arrive at the final goal of counting $n^L_{\rm dof}$, therefore, one has only to find mathematically 
a certain irreducible representation of $SO(D+1)$, which {\it properly} describes (as seen in what follows) 
the spatial quantities $( \hat{X}_u, R \hat{N})$ inside the bounded region with radius $L$, then one finds 
$n^L_{\rm dof}$ through counting the dimension of the representation. 

At this point, it is important to note that, as was remarked in advance in subsection 2.2, any generators of $SO(D+1)$ 
in YSTA  are defined by the differential operators on the $D-$dimensional unit sphere, $S^D$, i.e., 
${q_1}^2 + {q_2}^2 + \cdots + {q_{D-1}}^2 + {q_a}^2 + {q_b}^2 = 1,$ limiting its representations with $l$ to be 
integer.
   
On the other hand, it is well known that the irreducible representation of arbitrary high-dimensional $SO(D+1)$ 
on $S^D = SO(D+1)/SO(D)$ is derived in the algebraic way, $^{[12]}$ irrelevantly to any detailed knowledge of 
the decomposition equation (2.13), but solely in accord with the fact that $SO(D+1)$ in YSTA is defined originally 
on $S^D$, as mentioned above. One can choose, for instance, $SO(D)$ with generators $\hat{\Sigma}_{MN} (M,N=b, u)$, 
while $SO(D+1)$ with generators $\hat{\Sigma}_{MN}(M,N=a,b,u),$ where $u = 1,2,\cdots, D-1$. Then, it turns out that any irreducible representation 
of $SO(D+1)$, denoted by $\rho_l$, is uniquely designated by the maximal integer $l$ of eigenvalues of 
${\hat \Sigma}_{ab}$ in the representation, where ${\hat \Sigma}_{ab}$ is known to be a possible Cartan subalgebra 
of the so-called compact symmetric pair $(SO(D+1),SO(D))$ of rank $1$.$^{[12]}$ 

According to the so-called Weyl's dimension formula, the dimension of $\rho_l$ is given by$^{[12],[2],[8]}$
\begin{eqnarray}
 dim\ (\rho_l)= {(l+\nu) \over \nu} {(l+2\nu-1)! \over {l!(2\nu -1)!}},
\end{eqnarray}
where $ \nu \equiv (D-1)/2$ and $D \geq 2$.\footnote{This equation just gives the familiar result $dim\ (\rho_l)= 2l +1,$ 
in the case $SO(3)$ taking $D=2.$} 

Finally, we can find a certain irreducible representation of $SO(D+1)$ among those $\rho_l 's $  given above, 
which {\it properly} describes (or realizes) the spatial quantities inside the bounded region $V_D^L$. 
Now, let us choose tentatively $l = [L/\lambda]$ with $[L/\lambda]$ being the integer part of $L/\lambda$. In this case, 
one finds out that the representation $\rho_{[L/\lambda]}$ just {\it properly} describes all of generators of 
$SO(D+1)$ inside the above bounded spatial region $V_D^L$, because  $[L/\lambda ]$ indicates also the largest eigenvalue 
of any generators of $SO(D+1)$ in the representation $\rho_{[L/\lambda]}$ on account of its $SO(D+1)-$invariance and 
hence eigenvalues of spatial quantities $( \hat{X}_u, R \hat{N})$ are well confined inside the bounded region with 
radius {L}. As the result, one finds that the dimension of $\rho_{[L/\lambda]}$ just gives the number of spatial degrees 
of freedom inside $V_D^L$, $n^L_{\rm dof}$, when $[L/\lambda]\gg D$, as shown in (3.7).

\subsection{\normalsize KHR in the lower-dimensional spatial region $V_d^L$ }
According to the argument given for $V_D^L$ in the preceding subsection, let us study the 
kinematical holographic relation in the lower-dimensional bounded spatial region $V_d^L$ for the subsequent 
argument of the area-entropy relation in section 4. In fact, it will be given through a simple $D_0$ brane gas system 
formed inside $d\ (\leq{D-1})$-dimensional bounded spatial region, $V_d^L$, which is defined 
by
\begin{eqnarray}  
{X_1}^2 + {X_2}^2 + \cdots + {X_d}^2 = L^2,
\end{eqnarray}
instead of (3.5).

In this case, the boundary area of $V_d^L$, that is, ${\cal A}\ (V_d^L)$ is given by
\begin{eqnarray}
{\cal A}\ (V_d^L) =  ({\rm area\ of}\ S^{d-1}) ={(2 \pi)^{d/2} \over {(d-2)!!}} (L/\lambda)^{d-1} 
&&for\ d\ even
\nonumber\\                   =2 {(2\pi)^{(d-1)/2} \over {(d-2)!!}} (L/\lambda)^{d-1} &&for\ d\ odd, 
\end{eqnarray}
corresponding to Eq. (3.6).

On the other hand, the number of degrees of freedom of $V_d^L$, let us denote it $n_{dof} (V_d^L)$, is calculated 
by applying the arguments given for derivation of $n_{dof}^L$ in (3.7). In fact, it is found in a certain irreducible 
representation of $SO(d+1)$, a minimum subalgebra of YSTA, which includes the $d$ spatial quantities, $\hat{X}_1, 
\hat{X}_2, \cdots, \hat{X}_d$ needed to properly describe $V_d^L$, and is really constructed by the generators 
$\hat{\Sigma}_{MN}$ with $M,N$ ranging over $a,1,2, \cdots,d$. The representation of $SO(d+1)$, let us denote it 
$\rho_l\ (V_d^L)$ with suitable integer $l = [L/\lambda]$, is given on the representation space $S^d =SO(d+1)/SO(d)$, 
taking the subalgebra $SO(d)$, for instance, $\hat{\Sigma}_{MN}$ with $M,N$ ranging over 
$1,2,\cdots,d$, entirely in accord with the argument on the irreducible representation of $SO(D+1)$ given in the 
preceding subsection 3.1. 

One immediately finds that 
\begin{eqnarray}
&&n_{\rm dof}\ (V_d^L)  =  dim\ ( \rho_{[L/\lambda]}\ (V_d^L)) \sim {2 \over (d-1)!}{([L/\lambda]+ d-2)! \over ([L/\lambda]-1)!}
\nonumber\\
 &&\hskip3.5cm \sim {2 \over (d-1)!}  [L/\lambda]^{d-1}.
\end{eqnarray}
corresponding to (3.7), and there holds, from (3.10) and (3.11), the following kinematical holographic 
relation for $V_d^L$ in general
\begin{eqnarray}
\hspace{-3cm} [KHR] \hspace{2cm}       n_{\rm dof}\ (V_d^L)=
{\cal A}\ (V_d^L) / G_d,
\end{eqnarray}
with the proportional constant $G_d$
\begin{eqnarray}
G_d \sim {(2 \pi)^{d/2} \over 2}\ (d-1)!! &&for\ d\ even
\\
    \sim  (2 \pi)^{(d-1)/2}(d-1)!! &&for\ d\ odd,
\end{eqnarray}
corresponding to Eqs. (3.1)- (3.3) for $V_D^L$.  
     
\section{\normalsize Area-Entropy Relation in $D_0$ Brane Gas subject to YSTA}
  
\subsection{\normalsize $D_0$ Brane Gas Model in $V_d^L$ and Its Mass and Entropy}
Now, let us consider the central problem of the present paper, that is, the derivation of a possible area-entropy relation 
through a simple $D_0$ brane gas$^{[6]}$ model formed inside $V_d^L$ according to the idea of M-theory.$^{[7]}$ 
This implies that one has to deal with the dynamical system of the second-quantized $D_0$ brane field ${\hat D}_0$ 
inside $V_d^L$. In the present toy model of the $D_0$ brane gas, however, we avoid to enter into detail of the dynamics 
of $D_0$ brane system, but treat it as an ideal gas, only taking into consideration 
that the system is developed on $V_d^L$ subject to YSTA and its representation discussed above, but neglecting 
interactions of $D_0$ branes, for instance, with strings, as well as possible self-interactions among themselves. 

First of all, according to the argument given in the preceding subsection 3.2, the spatial structure of $V_d^L$ 
is described through the specific representation $ \rho_{[L/\lambda]}\ (V_d^L)$. Let us denote its orthogonal 
basis-vector system in Hilbert space I, as follows 
\begin{eqnarray}
 \rho_{[L/\lambda]}\ (V_d^L): \quad |\ m >,  \qquad  m= 1,2,\cdots, n_{\rm dof}(V_d^L).     
\end{eqnarray}   
 In the above expression, $n_{\rm dof}\ (V_d^L)$ denotes the dimension of the representation 
$\rho_{[L/\lambda]}\ (V_d^L)$, as defined in (3.11).

At this point, one should notice that the {\it second quantized} ${\hat D}_0$-brane field$^{[13]}$ on $V_d^L$ 
must be the linear operators operating on Hilbert space I, and described by $n_{\rm dof}(V_d^L) \times n_{\rm dof}(V_d^L)$ 
matrix under the representation $\rho_{[L/\lambda]}\ (V_d^L)$ like $< m\ |{\hat D}|\ n >$ on the one hand, and 
on the other hand each matrix element must be operators operating on Hilbert space II, playing the role of 
creation-annihilation of $D_0$ branes. On the analogy of the ordinary quantized local field, let us define 
those creation-annihilation operators through the diagonal parts in the following way:\footnote{On the other hand, 
the non-diagonal parts, $< m\ |{\hat D}|\ n >,$ 
are to be described in terms like ${\bf a}_m {\bf a}_n^\dagger$ or ${\bf a}_m^\dagger {\bf a}_n$ in accord with 
the idea of M-theory where they are conjectured to be concerned with the interactions between $[site\ m]$ and $[site\ n].$ 
The details must be left to the rigorous study of the second quantization of $D_0$-brane field.$^{[13]}$}    
\begin{eqnarray}
< m\ | {\hat D}|\ m >\  \sim\ {\bf a}_m\ {\rm or}\  {\bf a}_m^\dagger.
\end{eqnarray}
In the above expression, ${\bf a}_m$ and ${\bf a}_m^\dagger$, respectively, denote annihilation and creation 
operators of $D_0$ brane, satisfying the familiar commutation relations,
\begin{eqnarray}
&&[{\bf a}_m, {\bf a}_n^\dagger]= \delta_{mn},
\\
&&[{\bf a}_m, {\bf a}_n]= 0 .
\end{eqnarray}
 One notices that the labeling number $m$ of basis vectors, which ranges from $1$ to $n_{\rm dof}(V_d^L)$ 
plays the role of {\it spatial coordinates} of $V_d^L$ in the present noncommutative YSTA, corresponding to the so-called 
lattice point in the lattice theory. Let us denote the {\it point} hereafter $[site]$ or $[site\ m]$ of $V_d^L$.

Now, let us focus our attention on quantum states constructed dynamically in Hilbert space II by the creation-annihilation 
operators ${\bf a}_m$ and ${\bf a}_m^\dagger$ of $D_0$ branes introduced above at each [site] inside $V_d^L$. One should 
notice here the important fact that in the present simple $D_0$ brane gas model neglecting all interactions of $D_0$ branes, 
each $[site]$ can be regarded as independent quantum system and described in general by own statistical operator, while 
the total system of gas is described by their direct product. In fact, the statistical operator at each $[site\  m]$ 
denoted by ${\hat W}[m]$, is given in the following form,
\begin{eqnarray} 
{\hat W}[m] = \sum_ k  w_k\  |\ [m]: k >\  < k :[m]\ |,
\end{eqnarray}
with 
\begin{eqnarray}
 |\ [m]:  k  >  \equiv {1 \over \sqrt{k!}}({\bf a}_m^\dagger)^{k}|\ [m]:0 >.
\end{eqnarray}
That is, $|\ [m]:  k > ( k=0, 1, \cdots)$ describes the normalized quantum-mechanical state in Hilbert space II with 
$k$ $D_0$ branes constructed by ${\bf a}_m^\dagger$ on $|\ [m]:0 >,$ i.e. the vacuum state of $[site\ m]$.
\footnote{The proper vacuum state in Hilbert space II is to be expressed by their direct product.} And $w_k$'s denote 
the realization probability of state with occupation number $k$, satisfying $\sum_k w_k = 1.$ 

We assume here that the statistical operator at each $[site\ m]$ is common to every [site] in the present $D_0$ 
brane gas under equilibrium state, with the common values of $w_k$'s and the statistical operator of total 
system on $V_d^L$ , ${\hat W}(V_d^L)$, is given by  
\begin{eqnarray}
{\hat W}(V_d^L) = {\hat W}[1] \otimes {\hat W}[2] \cdots \otimes {\hat W}[m] \cdots \otimes {\hat W}[n_{dof}].
\end{eqnarray}

Consequently, one finds that the entropy of the total system, $S(V_d^L)$, is given by
\begin{eqnarray}
S(V_d^L) = - {\rm Tr}\ [{\hat W}(V_d^L)\ {\rm ln} {\hat W}(V_d^L)] = n_{dof}(V_d^L)\times S[site], 
\end{eqnarray}
where $S[site]$ denotes the entropy of each [site] assumed here to be common to every [site] and given by
\begin{eqnarray}
S[site] = - {\rm Tr}\ [ {\hat W}[site]\ {\rm ln}{\hat W}[site]] = - \sum_k w_k\  {\rm ln} w_k.
\end{eqnarray}

Comparing this result (4.8) with [KHR]\ (3.12) derived in the preceding section, we find an important fact that the 
entropy $S(V_d^L)$ is proportional to the surface area ${\cal A}\ (V_d^L)$, that is, a kind of area-entropy relation 
([AER]) of the present system:
\begin{eqnarray}
\hspace{-3cm} [AER] \hspace{2cm}         S(V_d^L) = {\cal A}\ (V_d^L)\ {S[site] \over G_d},
\end{eqnarray}
where $G_d$ is given by (3.13)-(3.14). 

Next, let us introduce the total energy or mass of the system, $M(V_d^L)$. If one denotes the average energy or mass 
of the individual $D_0$ brane inside $V_d^L$ by $\mu$, it may be given by
\begin{eqnarray}   
M(V_d^L) = \mu {\bar N}[site]\  n_{\rm dof}(V_d^L) \sim  \mu {\bar N}[site] {2  \over (d-1)!}  [L/\lambda]^{d-1},
\end{eqnarray}
where ${\bar N}[site]$ denotes the average occupation number of $D_0$ brane at each $[site]$ given by
\begin{eqnarray}
{\bar N}[site] \equiv \sum_k k w_k.
\end{eqnarray}

Comparing this expression (4.11) with (4.8) and (3.12), respectively, we obtain a kind of mass-entropy relation ([MER]) 
\begin{eqnarray}
\hspace{-2cm} [MER] \hspace{2cm}         M(V_d^L) / S(V_d^L) = \mu {\bar N}[site] / S[site],
\end{eqnarray}
and  a kind of area-mass relation ([AMR])
\begin{eqnarray}
\hspace{-3cm} [AMR] \hspace{2cm} M(V_d^L) = {\cal A}(V_d^L)\ {\mu {\bar N}[site] \over G_d}.
\end{eqnarray}
     
\subsection{\normalsize Schwarzschild Black Hole and Area-Entropy Relation In $D_0$ brane Gas System}

In the preceding subsection 4.1, we have studied $D_0$ brane gas system and derived area-entropy relation $[AER]$ 
(4.10), mass-entropy and area-mass relations, $[MER]$ (4.13) and $[AMR]$ (4.14), which are essentially based on 
the kinematical holographic relation in YSTA studied in section 3. 

At this point, it is quite important to notice that these three relations explicitly depend on 
the following ${\it static}$ quantities of the gas system, $\mu$, ${\bar N}[site]$ and $S[site]$, that is, the average 
energy of individual $D_0$ brane, the average occupation number of $D_0$ branes and the entropy at each [site], 
which are assumed to be common to every [site], while these quantities turn out to play an important role in 
arriving finally at the area-entropy relation in connection with black holes, as will be seen below.   

Now, let us investigate how the present gas system tends to a black hole. We assume for simplicity that the system 
is under $d=3$, and becomes a Schwarzschild black hole, in which the above quantities acquire certain limiting 
values, $ \mu_S$, ${\bar N}_S[site]$ and $S_S [site]$, while the size of the system, $L$, becomes $R_S$, 
that is, the so-called Schwarzschild radius given by 
\begin{eqnarray}   
R_S =  2 G M(V_3^{R_S})/c^2,
\end{eqnarray}
where $G$ and $c$ denote Newton 's constant and the light velocity, respectively, and $M(V_3^{R_S})$ is given by 
Eq. (4.11) with $L= R_S$, $\mu = \mu_S$ and ${\bar N}[site] = {\bar N}_S[site]$. Indeed, inserting the 
above values into Eq.(4.11), we arrive at the important relation, called hereafter the black hole condition [BHC], 
\begin{eqnarray}
\hspace{-1cm} [BHC] \hspace{2cm} M(V_3^{R_S}) = {\lambda^2 \over 4\mu_S {\bar N}_S [site]}{c^4 \over G^2} 
= {M_P^2 \over 4 \mu_S {\bar N}_S[site]}.
\end{eqnarray}
In the last expression, we assumed that $\lambda$, i.e., the small scale parameter in YSTA is equal  
to Planck length $l_P = [G \hbar / c^3]^{1/2} = \hbar /( c M_P )$, where $M_P$  denotes Planck mass.

On the other hand, we simply obtain the area-entropy relation [AER] under the Schwarzschild black hole by 
inserting the above limiting values into [AER] (4.10)   
\begin{eqnarray}
S(V_3^{R_S}) = {\cal A}\ (V_3^{R_S})\ {S_S[site] \over 4 \pi},
\end{eqnarray}
noticing that $G_{d = 3} =4 \pi.$

At this point, one finds that it is a very important problem how to relate the above area-entropy relation 
under a Schwarzschild black hole with [AER] (4.10) of $D_0$ brane gas system in general, which is 
derived irrelevantly of the detail whether the system is a black hole or not. As was mentioned in the beginning 
of this subsection, however, the problem seems to exceed the applicability limit of the present toy model 
of $D_0$ brane gas, where the system is treated solely as a {\it static} state under {\it given} values of 
parameters, $\mu$, ${\bar N}[site]$ and $S[site]$, while the critical behavior around the formation of Schwarzschild 
black hole must be hidden in a possible ${\it dynamical}$ change of their values. 

In order to supplement such a defect of the present static toy model, let us try here a Gedanken-experiment, 
in which one increases the entropy of the gas system $S(V_3^L)$, keeping its size $L$ at the initial value $L_0$, 
until the system tends to a Schwarzschild black hole, where Eqs. (4.16) and (4.17) with $R_S = L_0$ hold. Then, one 
finds that according to [AER] (4.10), the entropy of $[site]$, $S[site]$ increases proportionally to $S(V_3^L)$ and 
reaches the limiting value $S_S [site]$, starting from any initial value $S_0[site]$ prior to formation of the black 
hole, because ${\cal A}(V_d^L)$ in Eq.(4.10) is invariant during the process. Namely, one finds a very simple fact 
that $S_0[site] \leq S_S [site].$  However, this simple fact combined with [AER] (4.10) leads us to the following 
form of a new area-entropy relation which holds throughout for the $D_0$ brane gas system up to the formation of 
Schwarzschild black hole,\footnote{Similarly, by the second Gedanken-experiment, in which one increases 
the total mass of gas system $M(V_3^L)$ with the fixed size $L_0$ in connection with [AMR] (4.14), 
in place of the increase of the entropy of gas system $S(V_3^L)$ in the first Gedanken-experiment, 
one obtains a new area-mass relation [AMR], $M(V_3^L) \leq {\cal A}(V_3^L) \mu_S {\bar N}_S[site] / 4 \pi$.}
\begin{eqnarray}
\hspace{-3cm} [AER] \hspace{2cm}  S(V_3^L) \leq {{\cal A} (V_3^L) S_S[site] \over 4 \pi},
\end{eqnarray}
where the equality holds for Schwarzschild black hole, as seen in Eq. (4.17).

\subsection{\normalsize The Possible Inner Relation between Our Approach and Bekenstein-Hawking Relation}

We have derived the area-entropy relation [AER] (4.18) together with (4.10) in our toy model of $D_0$ brane gas subject to Yang's quantized 
space-time algebra, YSTA. Indeed, it is essentially based on the fundamental nature of the noncommutative geometry
of YSTA, that is, the kinematical reduction of spatial degrees of freedom and holographic relation in YSTA, 
which was pointed out in ref. [2]and extended to the lower dimensional region,$^{[1]}$ $V_d^L$, as shown by [KHR] (3.12) 
in section 3. The relation [AER] (4.18) is to be compared with the original Bekenstein proposal
\begin{eqnarray}
S \leq \eta {\cal A},
\end{eqnarray}
where the proportional constant $\eta$ is now given in terms of a physical quantity, i.e., the partial entropy of 
the individual [site] of ${D_0}$ brane gas system under Schwarzschild black hole, that is, $S_S[site]$, as follows, 
\begin{eqnarray}
\eta = {S_S[site] \over 4 \pi}.
\end{eqnarray}

In addition, it is well-known that the Bekenstein proposal (4.19) was extended to the Bekenstein-Hawking Area-entropy 
relation 
\begin{eqnarray}
S \leq {\cal A} /4
\end{eqnarray}
with $\eta$ fixed to be $1/4$ through the investigation of the so-called Hawking radiation of black hole, 
which suggests to us more specifically that
\begin{eqnarray}
 S_S[site] = \pi.
\end{eqnarray}

In order to make clear the implication of the above {\it constraint} (4.22), let us supplement our preceding 
arguments by introducing the concept of temperature $T$ of the present gas system of $D_0$ branes, through the entropy 
of individual [site] mentioned above. It was assumed to be common to every $[site]$ of $D_0$ brane gas system in 
equilibrium, given by $S[site] =  {\rm Tr}\ [ {\hat W}[site]\ {\rm ln}{\hat W}[site]]= - \sum_k w_k\  {\rm ln} w_k $ 
in Eq. (4.9). 

Now let us take the following familiar expression for $w_k$'s,
\begin{eqnarray}
 w_k = e^{-\mu k/T} / Z(T)
\end{eqnarray}
where
\begin{eqnarray}
Z(T) \equiv \sum_{k=0}^\infty e^{-\mu k/T} = 1 / (1- e^{-\mu / T}). 
\end{eqnarray}

Then, one finally finds that 
\begin{eqnarray}
S[site]\equiv - \sum_k w_k\  {\rm \ln} w_k = - \ln (1 - e^{- \mu /T}) + {\mu \over T}\ ( e^{\mu /T}- 1)^{-1},
\end{eqnarray}
\begin{eqnarray}
{\bar N}[site] \equiv \sum_k k w_k = ( e^{\mu / T}- 1)^{-1}.
\end{eqnarray}
and there holds the relation
\begin{eqnarray}
{d \over dT} {(\mu \bar N[site])} =  T\  {d \over dT} S[site],
\end{eqnarray}
assuming that $\mu$ is independent of $T$.

At this point, let us apply the above result to our gas system under Schwarzschild black hole condition 
[BHC] considered in the preceding subsection 4.2, where all physical quantities were denoted with subscript $S$, 
such like $T_S$. First of all, one notices that Eq. (4.26) combined with Eq. (4.16) [BHC] 
\begin{eqnarray}
{\bar N}_S[site] (= ( e^{\mu_S /T_S}-1)^{-1}) = 1 / (4 \mu_S M_S)
\end{eqnarray}
gives rise to the important relation which relates between temperature $T_S$ and mass $M_S$ under [BHC] through 
the intermediation of $D_0$ brane mass $\mu_S$, that is,  
\begin{eqnarray}
 (e^{\mu_S / T_S} - 1) =  4  \mu_S M_S  
\nonumber
\end{eqnarray}
or
\begin{eqnarray}
 T_S  = \mu_S / \ln (1 + 4 \mu_S M_S).
\end{eqnarray} 
where Planck units in $D=4$ are used, with $M_P = l_P = \hbar = c = k =1$.

Furthermore, from Eq. (4.25), one finds
\begin{eqnarray}
S_S [site]\ ( = - \ln (1 - e^{- \mu_S /T_S}) + {\mu_S \over T_S}\ ( e^{\mu_S /T_S}- 1)^{-1})
\nonumber\\ =(1+ {1 \over {4 \mu_S M_S}}) \ln (1 + 4\mu_S M_S) - \ln (4 \mu_S M_S)
\nonumber\\ = \ln (1+ (4\mu_SM_S)^{-1}) + {1 \over {4 \mu_S M_S}} \ln (1 + 4\mu_S M_S) 
\end{eqnarray}

On the bases of the above relations, let us try in our present scheme to find the so-called 
Hawking radiation temperature $T_{H.R.}$ of the gas system under Schwarzschild black hole, which is defined by 
\begin{eqnarray}
T_{H.R.}^{-1} =  dS_S/dM_S.
\end{eqnarray}
Noticing the relation [MER] (4.17) with $ {\cal A} = 4 \pi R_S^2 =16 \pi M_S^2$, we immediately 
find 
\begin{eqnarray} 
T_{H.R.}^{-1} ={d \over dM_S} S_S  = {d \over d M_S} (16 \pi M_S^2 S_S[site]/{4 \pi})     \cr
\nonumber\\             = 8 M_S S_S[site] + 4M_S^2  {d \over d M_S}{S_S[site]}.
\end{eqnarray}

At this point, it is quite important in the calculation of the second term of the last 
expression of (4.32), how to consider the possible dependence of $\mu_S$ on the total mass $M_S$ 
in $S_S[site]$, whose explicit expression is given in (4.30). In what follows, let us first 
two simple exteme cases, A) and B). 

In the case A), let us assume that $\mu_S$ is independent of $M_S$ according to the case of  
derivation of the relation (4.27), in which we assumed simply $\mu$ to be independent of $T$. 
Then we obtain
\begin{eqnarray}
T_{H.R.}^{-1} &= 8 M_S \ln (1 + (4\mu_S M_S)^{-1} ) +  ( 1 / \mu_S )  \ln (1 + 4 \mu_S M_S) \cr
\nonumber\\  &= 8 M_S \ln (1 + (4\mu_S M_S)^{-1} ) + T_S^{-1}.   \qquad Case A)
\end{eqnarray}

In the case B), let us assume that $\mu_S M_S ( = ( 4 \bar N_S[site])^{-1},$ see (4.28)) is some fixed constant, 
as in the case of Bekenstein-Hawking constraint (4.22) on $S_S [site]$ (4.30). Then we simply obtain
\begin{eqnarray}
T_{H.R.}^{-1} = 8 M_S S_S[site] \hspace{8cm}
\nonumber \\   = 8 M_S \ \ln (1+ (4\mu_SM_S)^{-1}) + {2 \over { \mu_S}} \ln (1 + 4\mu_S M_S).   
\quad Case B)
\end{eqnarray}

Finally, it is quite important to notice that if we take seriously, from the beginning, the Bekenstein-Hawking 
relation
\begin{eqnarray}
S_S = {\cal A} /4,
\end{eqnarray}
as the first principle, it compels us to take the stringent constraint (4.22), that is, 
\begin{eqnarray}
S_S[site] = \ln (1+ (4\mu_SM_S)^{-1}) + {1 \over {4 \mu_S M_S}} \ln (1 + 4\mu_S M_S)) = \pi,
\end{eqnarray} 
which means that $\mu_S M_S $ or $\bar N_S[site] (= 1/(4\mu_S M_S)$, i.e., the average occupation number of $D_0$ 
branes at each [site] must take  a certain fixed values, that is, $\mu_S M_S (/M_P^2) \sim 0.03, \
\bar N_S[site] \sim 1/ 0.12$, irrespectively of individual black holes and at the same time leads us, through (4.34), 
to the well-known Hawking radiation temperature,
\begin{eqnarray}
T_{H.R.} (= 1/(8 M_S S_S[site])) = {1 \over {8 \pi M_S}}.    
\end{eqnarray}
One should notice here the similarity between $\mu_S\sim 0.03/M_S$ and $T_{H.R.} (=1/(8\pi M_S)) \sim 0.04 /M_S.$
 
We will further intensify the above arguments in the next section.

\section{\normalsize Discussion and Concluding Remarks}

 In the present paper, we have started with the argument of the kinematical holographic relation [KHR] (1.1) 
found in the Lorentz-covariant Yang's quantized space-time algebra (YSTA). As was emphasized 
there, it essentially reflects the fundamental nature of the noncommutative geometry of YSTA itself, that is, a definite 
kinematical reduction of spatial degrees of freedom in comparison with the ordinary lattice space. Furthermore, 
YSTA, which is intrinsically equipped with short- and long-scale parameters, $\lambda (=l_P)$ and $R$, gives a finite 
number of spatial degrees of freedom for any finite spatial region and provides a basis for the field theory 
free from ultraviolet- and infrared-divergences,$^{[2],[8],[14]}$ Therefore, the argument on the holographic 
relation and area-entropy relation of $D_0$ brane gas system extended in the present paper has aimed to 
respect {\it the first principle} of YSTA as much as possible, although it may be too crude and simple 
to treat physics around Planckian scale. 

From this point of view, the present author ought to give importance, for instance, to the expression of 
area-entropy relation [AER] (4.18) or (4.19) with $\eta = S_S[site]/4 \pi$ (4.20). The former relation (4.18), which has 
been naturally derived from the kinematical holographic relation mentioned above and extensively studied in 
subsection 4.3 in connection with 
Bekenstein-Hawking relation and Hawking radiation temperature $T_{H.R.}$. We expect that it possibly gives us a clue 
to search for physics around the Planckian scale, from the standpoint that the Bekenstein-Hawking relation as well as 
Hawking radiation temperature provides us, not {\it a priori doctrine} of black holes, but rather important empirical 
knowledge, as stated in connection with (4.36) at the end of the preceding subsection 4.3. From this point of view, 
we further consider on several results derived in the preceding section 4, in particular, the subsection 4.3.

1. It is important first to focus our attention on $S_S[site]$, on which a certain kind of universality 
was pointed out, such like $S_S[site] = \pi\ (4.22)$ in connection with Bekenstein-Hawking relation. Let us start with 
Eq. (4.17), which was given prior 
to [AER] (4.18):
\begin{eqnarray}
S_S(V_3^{R_S}) = {\cal A}\ (V_3^{R_S})\ {S_S[site] \over 4 \pi},  
\end{eqnarray}
where $S_S(V_3^{R_S})$ on the left-hand side is given by (4.8);
\begin{eqnarray}
S_S(V_3^{R_S}) = n_{dof}(V_3^{R_S}) S_S[site],   
\end{eqnarray}
with $R_S = 2M(V_3^{R_S})$ (4.15).

Now, let us take an extreme case, $ R_S = \lambda\ ( = l_P)$ in (5.1) and (5.2). Then, taking into consideration that 
$ {\cal A} (V_3^{\lambda}) = 4 \pi$ and $n_{dof}(V_3^\lambda) = 1$ according to (3.10) and (3.11), respectively, 
one finds that both Eqs. (5.1) and (5.2) lead us to the important result,
\begin{eqnarray}
S_S(V_3^{R_S=l_P}) = S_S[site].
\end{eqnarray}

This result  tells us that $S_S[site]$ under consideration is nothing but the extreme entropy of the {\it elementary} 
black hole, $S_S(V_3^{R_S=l_P})$, which consists of a single [site] with Planckian scale and must have the mass 
$M_S(V_3^{R_S=l_P}) (= R_S /2$, see (4.15)) $= l_P/2$ or $M_P/2$ in full units. It should be noted that this conspicuous 
statement is in accord with the result derived from the Bekenstein-Hawking relation (4.35) itself, which also 
tells us that $S_S $ becomes $\pi$ when $R_S =l_P (= 1)$ and ${\cal A} =4 \pi$ in (4.35) and leads us to 
$S_S[site] = \pi$. (4.36).

2.  Let us turn our attention to ${\bar N}_S[site] (= 1/(4\mu_S M_S)),$ which appeared in (4.28) and was pointed out 
to have a certain universal nature as $S_S[site]$ mentioned above. In fact, according to the argument on 
$S_S[site]$ given above, one finds out that ${\bar N}_S[site]$ is nothing but the average number of $D_0$ brane inside 
the {\it elementary} black hole mentioned above, and its value is constrained to have a certain fixed value, that is, 
\begin{eqnarray}
{\bar N}_S[site](= 1/(4\mu_S M_S), see (4.28)) \sim 1/0.12,
\end{eqnarray} 
in accord with $S_S [site] = \pi $, as was stated after Eq. (4.36).

3. Next, let us investigate the Schwarzschild black holes with $R_S$ in general other than the elementary black 
hole considered above. 

With respect to the entropy, Eq. (5.2) clearly tells  us that the entropy of the {\it elementary} black hole, 
$S_S[site]$, plays the role of the {\it element} of the entropy of black holes in general. 

With respect to the mass of black holes in general, let us remember Eq. (4.14), which becomes 
\begin{eqnarray}
M_S (= M(V_3^{R_S})) = {\cal A}(V_3^{R_S})\ {\mu_S {\bar N}_S[site] \over 4 \pi} 
( = n_{dof}(V_3^{R_S})\ \mu_S {\bar N}_S[site]),
\end{eqnarray}
when the system tends to schwarzschild black hole with radius $R_S.$
       
 One notices that Eq. (5.5) combined with Eq. (5.4) or (4.28), nicely reproduces the relation (4.15), 
\begin{eqnarray}
R_S = 2 M_S,
\end{eqnarray}
as expected, on account of $ {\cal A}(V_3^{R_S}) = 4\pi R_S^2. $

Eq. (5.4) further gives an interesting result
\begin{eqnarray}
\mu_S \sim 0.03/ M_S = 0.06/ R_S. 
\end{eqnarray}
This result tells us very importantly that the effective mass of $D_0$ brane inside black holes, $\mu_S$, is inversely 
proportional to the respective black hole mass $M_S$, almost in accord with the Hawking radiation temperature, 
$ T_{H.R.}= 1/(8\pi M_S) \sim 0.04 / M_S,$ as was remarked at the end of subsection 4.3.

4. Finally, motivated by the last statement in 3., let us introduce a new {\it ansatz} as the first principle in our 
scheme, which enables us to equate $\mu_S$ self-consistently to $T_{H.R.}$ defined originally in Eq. (4.31). 
As will be easily 
understood from the argument in 3., this possibility turns out to be only possible when the parameter $\eta$ 
in Eqs. (4.19) or (4.20), is treated as a free parameter, not necessarily fixed to be $1/4$ from the beginning, 
but rather to be decided self-consistently.

As a matter of fact, one finds out that the new ansatz is well brought in our present scheme, together with the relation 
(4.20):
\begin{eqnarray}
S_S[site] = 4 \pi \eta,
\end{eqnarray}
\begin{eqnarray}
[Ansatz]  \hspace{1cm} \mu_S = T_{H.R.} = {1 \over (8 \pi M_S)(4 \eta)}, 
\end{eqnarray}
taking into account $ T_{H.R}^{-1} = 8 M_S S_S[site]$ (4.34).

Noticing the relation 
\begin{eqnarray}
4\mu_S M_S (= {\bar N}_S [site]^{-1}) = (8\pi \eta)^{-1},
\end{eqnarray}
derived from (5.9), one finds out that the parameter $\eta$ is determined from the equation 
\begin{eqnarray}
\ln(1+ (8\pi \eta)) + 8\pi \eta \ln (1 + (8\pi \eta)^{-1}) ( = S_S[site], see (4.30))= 4\pi \eta,
\end{eqnarray}
with the result
\begin{eqnarray}
\eta \sim 0.22\ ( \sim 1/4  \times 0.88).
\end{eqnarray}
On the other hand, Eq. (5.9) gives rise to 
\begin{eqnarray}
\mu_S = T_{H.R.} = {1 \over (8 \pi M_S)} \times (0.88)^{-1}.
\end{eqnarray}

Needless to say, we are not granted to much emphasize the physical significance of the deviation factor $0.88$ in 
Eqs.(5.12) or (5.13) with respect to $\eta$ or $T_{H.R.}$, if we consider our simple and crude method to treat such 
a $D_0$ brane gas system under $[BHC]$.   
 
In conclusion, it should be emphasized again that almost all results derived in this paper, essentially reflect 
the kinematical holographic relation (3.1) or (3.12) in YSTA.  So, it is interesting to examine how the 
kinematical reduction of spatial degrees of freedom expected in the noncommutative space-time algebra holds 
in general besides YSTA extensively studied so far. This consideration will give us an important clue 
to seek for a candidate for the ultimate theory, which is expected to satisfy the kinematical holographic relation 
and to be free from UV- and IR-divergences.$^{[2],[14]}$

\end{document}